# Repair of magnetism in oxidized graphene nanoribbons

D.W. Boukhvalov

*Computational Materials Science Center, National Institute for Materials Science,*

*1-2-1 Sengen, Tsukuba, Ibaraki 305-0047, Japan*

*Novel route for the establishing of magnetism in realistic oxidized graphene nanoribbons is proposed. Modelling of the migration of hydroxyl groups from central part to the zig-zag edges of graphene nanoribbons passivated by oxygen are performed with using density functional theory. The presence of hydroxyl groups near the edges leads formation of dangling bonds there instead saturated due to oxidation and repairs of magnetism diminished by edges oxidation. The route of manufacturing and stability of new type of magnetic graphene nanoribbons are also discussed.*

E-mail: D.Bukhvalov@science.ru.nl

## 1. Introduction

Since the first prediction of remarkable properties of graphene nanoribbons (GNR) [1, 2] they are in focus of modern theoretical [3-9] experimental [10-16] nanoscience. The most intriguing magnetic properties of GNR are associated with zig-zag edges. Unpaired electrons there are not only cause of the magnetism [3, 4] but also source for enormous chemical activity [17]. Unfortunately recent methods for GNR manufacturing provide unreducible edges oxidation [16] and diminishment of magnetism there due to dangling bonds saturation [17]. In absence of oxidized species another problem arises. Magnetic zig-zag edges demonstrate reconstruction to nonmagnetic pentagon-heptagon structure predicted theoretically [7-9] and later confirmed experimentally [15, 16]. From the recent experimental and theoretical data, it can be concluded that edges functionalization prevent reconstruction but [7], anyway, provides vanishing of magnetism there [17].

One of the possible ways of construction of magnetic graphene nanoribbons is partial hydrogenation of graphene sheets [18]. This route is quite promising but width of these "ribbons" inside graphene sheets is difficult to control. In the current work, I have reported the alternative way for repair of magnetism in oxidized graphene nanoribbons.

## 2. Graphene oxide nanoribbons

Graphene oxide (GO) now is one of the main compound for graphene production [19]. In recent works the high level of GO purity has been achieved (see for review Refs. [20-22]). The main problem of GO total reduction is the colossal binding energy between carbon and hydroxyl groups [23]. Maximal obtained carbon to oxygen ratios for the strongly reduced GO samples about 10:1. The GO could be also the source for GRR production by oxidative and non-oxidative cut.

Partial oxidation near-edger regions in GO nanoribbons are provided by the migration of hydroxyl groups from central part to the edges (see Fig. 1). Limited amount of hydroxyl groups in GO indicates possibility of formation of partially oxidized graphene nanoribbons with final width of oxidized part on the edges. The main challenge for studying these compounds is the energetics of formation and stability of magnetic configurations.

## 3. Computational method and model

Density functional theory (DFT) calculations had been carried out with the same pseudopotential code SIESTA [24] used for our previous modeling of GNR [17] and GO [23]. All calculations are done with using generalized gradient approximation (GGA-PBE) [25] which is suitable for description of graphene-adatom chemical bonds [26]. For carefully modeling of oxidized GNR I used a graphene ribbon containing four row 16 carbon atoms each (see Fig. 1a). The width of the ribbon is 15.7 Å. Thus size of supercell, it is guaranteed absence of any overlap between distorted areas around chemisorbed species along the ribbon groups [26]. All calculations were carried out for energy mesh cut-off 360 Ry and k-point mesh 2×8×1 in Mokhorst-Park scheme [27]. During the optimization, the electronic ground state was found self-consistently using norm-conserving pseudo-potentials for cores and a double-$\zeta$ plus polarization basis of localized orbitals for carbon and oxygen, and double-$\zeta$ basis for hydrogen. Optimization of the bond lengths and total energies was performed with an accuracy of 0.04 eV/Å and 1 meV, respectively. When drawing the pictures of density of states, a smearing of 0.2 eV was used.

## 4. Results and discussions

For the checking of the energies required for hydroxyl group migration from the center of GNR (Fig. 2a) to the edge (Fig. 2b) calculations of the total energies for each intermediate step have been performed. Results of calculation are presented on Fig. 1c. For the case of hydroxyl groups migration in contrast with hydrogen [18] no energetically favorable intermediate positions have been observed. In agreement with recent calculations of migration barriers [28] for studied concentration of these groups barrier of migration between equivalent positions should be about 0.3 eV. For estimate realistic values of migration barriers to the energy difference between steps of migration noted value should be added. All steps of migration except last are endothermic. The final configuration is more energetically favorable than initial and much more energetically favorable than previous step. Reported results evidence that hydroxyl groups could be moved from the central part of GO to the edges and this process is irreversible at ambient conditions. Next step of the studies require the modeling of similar migration process from the center to the borders of GNR in presence there hydroxyl groups. From results of calculations presented on Fig. 3 we can conclude that the presence of several hydroxyl groups near the borders facilitates further migration of next hydroxyl groups to the borders. The cause of found diminution of migration barrier is the significant distortions of the nanoribbons due to the placement hydroxyl groups in the center and on the edges (sees Fig. 3a). For check the role of graphene flat distortions calculations for the same migration of almost two times wider ribbon had been performed. Reduce of corrugation in wider GNR provides higher energies required for hydroxyl group migration. It needs to note that the highest energy corresponding to the penultimate step of the migration is weakly depends from the width of the ribbons. Hence the width of the GNR plays no significant role for the energetics of hydroxyl group migration and stability of magnetism in discussed systems.

Final configuration of oxidized GNG with next-border states fully occupied by hydroxyl groups are shown on Fig. 1b. The total energy of hydroxyl group for this configuration is 0.66 eV smaller that for the single pair of hydroxyl groups in central part of GNR (Fig. 2a). Electronic structure of GNR is dramatically changed due to the appearance of the hydroxyl groups near the border. For the pure GNR with oxidized edges electronic structure correspond with non-magnetic metal, and building the line of hydroxyl groups restore the magnetism on the borders with magnetic moment 0.25 $\mu_B$ per carbon atom on the edge. The source for repair of magnetism on the edges is new dangling bonds which come out after hydroxyl groups migrations (see Fig. 4). The energy difference between ferromagnetic and anfferromagnetic configuration across the ribbon is 3 meV per zig-zag step ($J_a$ on Fig. 4c), and 15 meV per zig-zag step between ferromagnetic and anfferromagnetic configurations along the ribbon ($J_b$ on Fig. 4c). Calculated values for magnetic interactions suggest the possibility of predicting magnetic effects [1-4] in studied compounds. Formation of midgap states [29-31] enhances magnetic interactions there and makes studied compounds very attractive for various applications. Reported results provide the explanation for the appearance of magnetism in GO samples after annealing [32].

Final check of the stability of studied new magnetic GNR are required.

Magnetism based on the dangling bonds could be destroyed due to saturation of these bonds [18, 26]. This process could take place for the further migration of next hydroxyl groups from inside of GNR. Results of this modelling are presented on Fig. 5. The energies required for the migration of the next hydroxyl groups are higher than for the migration of same groups to the non-functionalized edges (Fig. 3) and penultimate step of migration is less energetically favourable than last one. The limited numbers of these groups in reduced GO also prevent this process. Another way for diminution of magnetism on the oxidized edges is moving part of hydroxyl groups back. The energy difference between initial and final configuration for single hydroxyl group (Fig. 6) is 2.71 eV, it make this migrations very energetically unfavourable.

## 5. Conclusions

The energetics of migration of hydroxyl groups from the central part of GO nanoribbons to the edges have been examined by DFT modelling. It is shown that presence of –OH groups near the realistic zig-zag edges passivated by oxygen provide formation of new dangling bonds instead of saturated ones due to oxidation. Proposed configurations are rather stable and correspond with repair of the magnetism on the edges. Reported novel type of magnetic GNR could be realized by oxidative cut of GNR from reduced GO with further annealing. Reported fluidity of hydroxyl groups to the edges could be also used for production of pure graphene form reduced GO by annealing of samples and further non-oxidative cut (for example by argon plasma [14]) of edges with all hydroxyl groups and oxygen atoms.

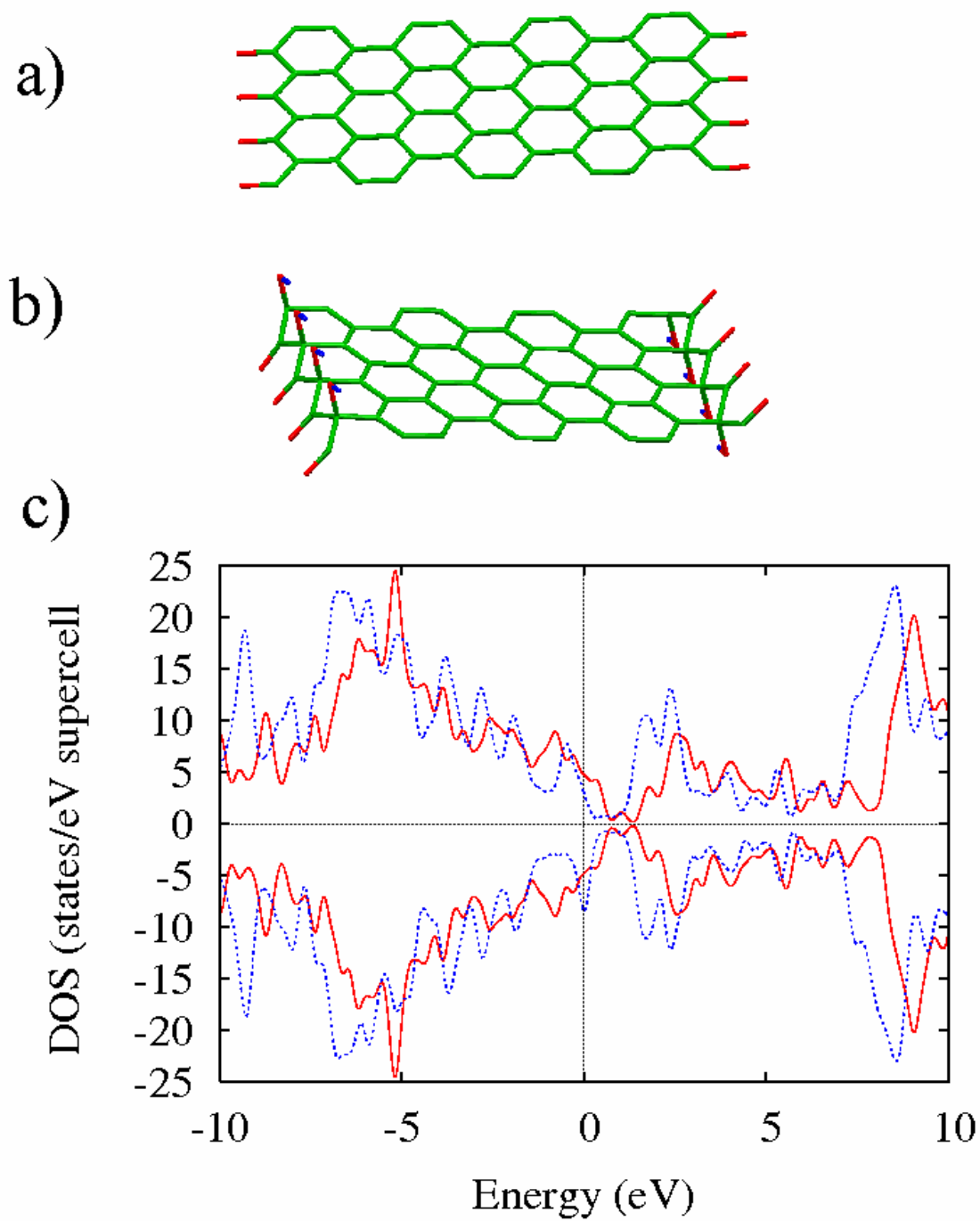

**Figure 1.** Optimized atomic structures for oxidized GNR with zig-zag edges (a) and oxidized GNR functionalized with hydroxyl groups (b). Carbon atoms are shown by green, oxygen by red and hydrogen by blue. Densities of states for compounds on panels (a) – solid red line, and panel (b) – dashed blue line.

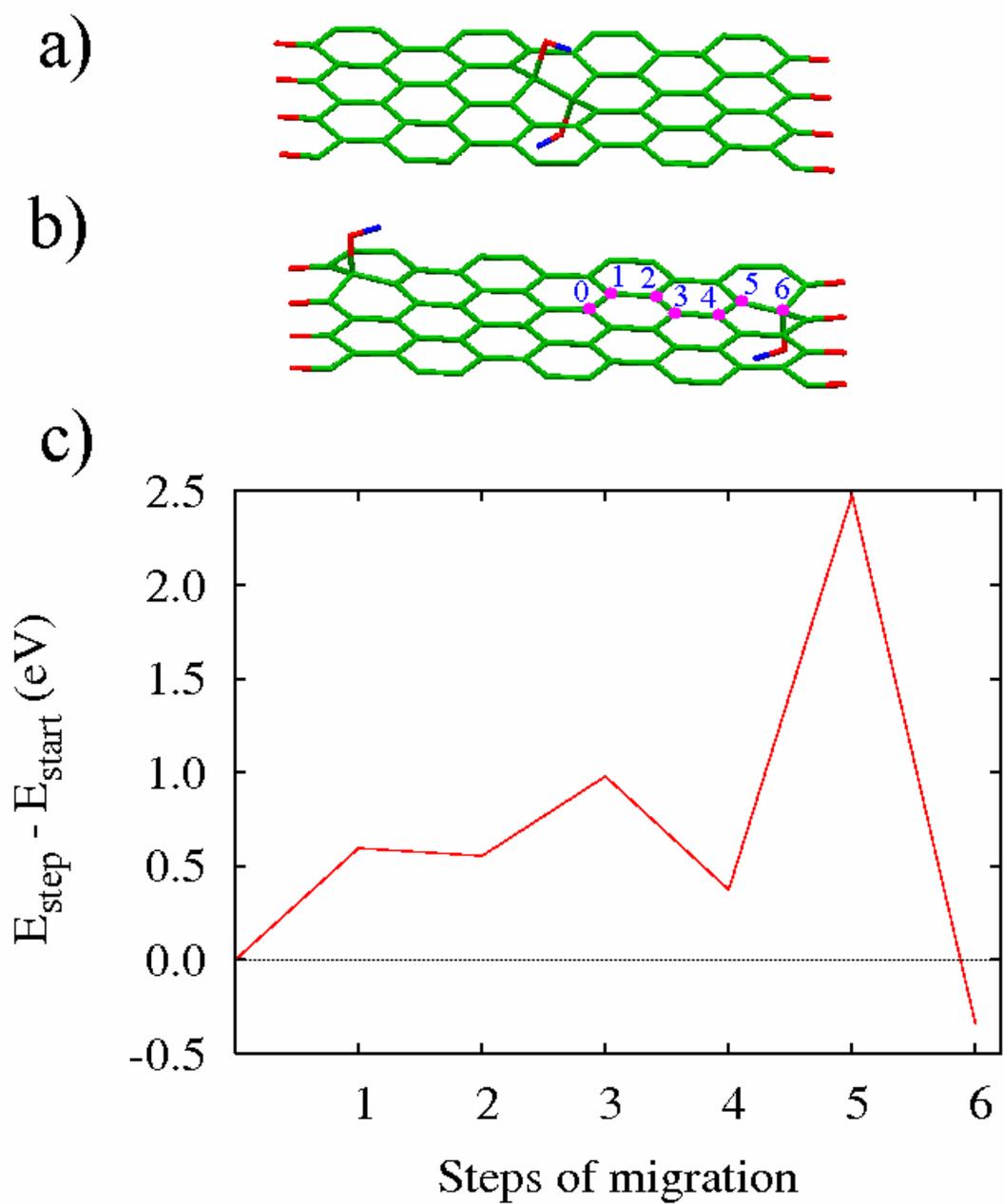

**Figure 2.** Optimized atomic structure for initial (a) and final (b) steps of the migration of hydroxyl groups from the centre to the edges of oxidized GNR. The energy differences between current and initial steps of migration are shown on panel (c). The steps of migration are noted on panel (b).

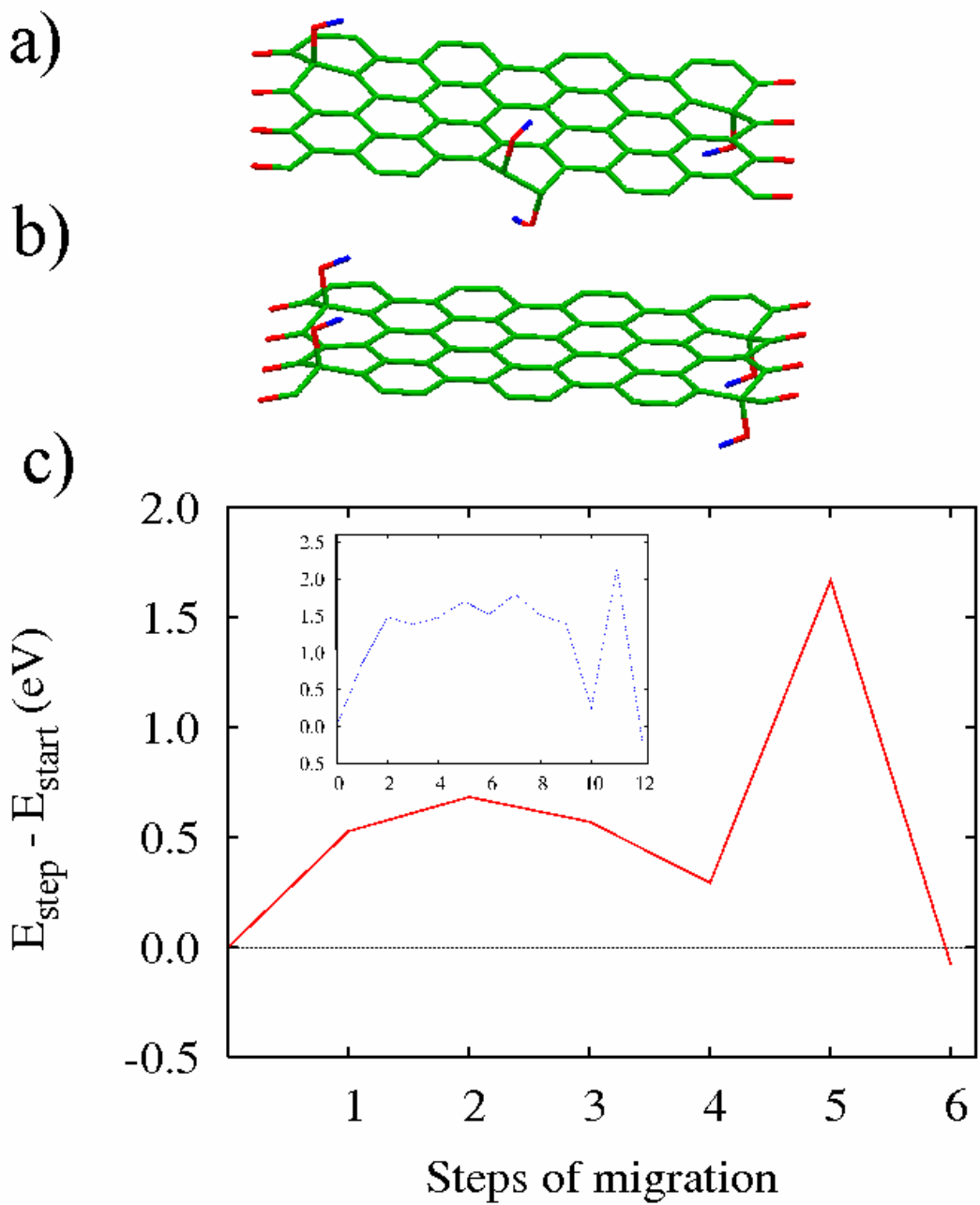

**Figure 3.** Optimized atomic structure for initial (a) and final (b) steps of the migration of hydroxyl groups from the centre to the edges of oxidized GNR functionalized with single hydroxyl groups. The energy differences between current and initial steps of migration are shown on panel (c). On inset of panel (c) the same energy differences for similar migration process in nanoribbon of 28.5 Å widths.

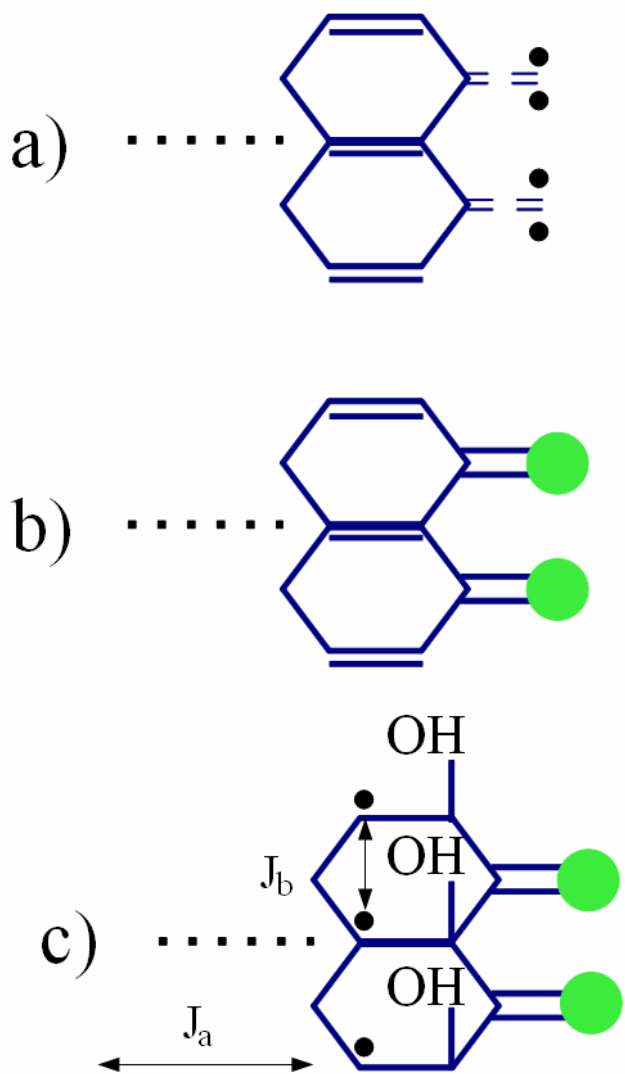

**Figure 4.** A sketch of the chemical bonds of unpassivated zig-zag edge of graphene (a), oxidized zig-zag edge (b), and oxidized zig-zag edge functionalized with hydroxyl groups (c). Oxygen atoms are shown by big green circles, unpaired electrons on dangling bonds by small black circles.

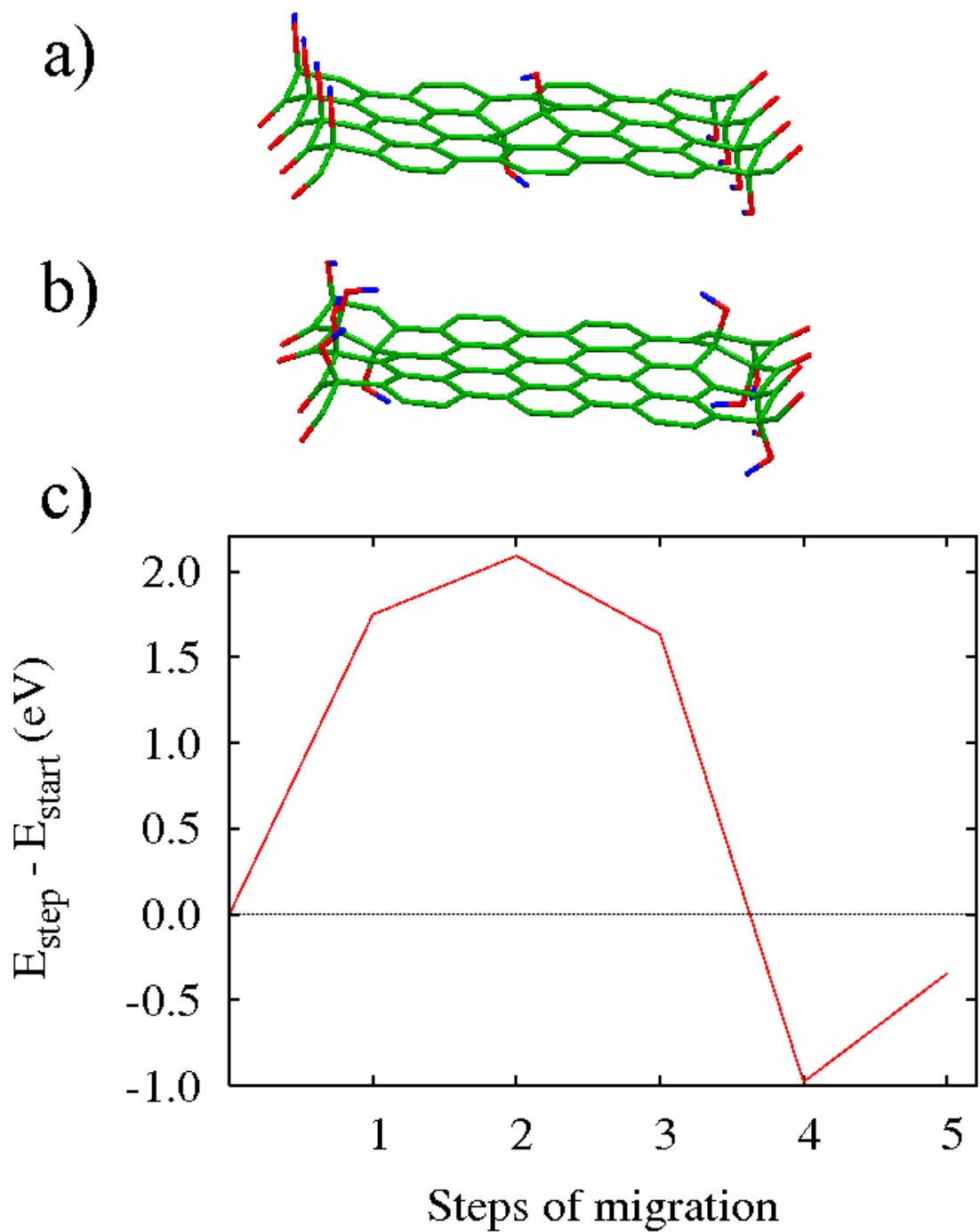

**Figure 5.** Optimized atomic structure for initial (a) and final (b) steps of the migration of hydroxyl groups from the centre to the edges of oxidized GNR functionalized with the lines of hydroxyl groups. The energy differences between current and initial steps of migration are shown on panel (c).

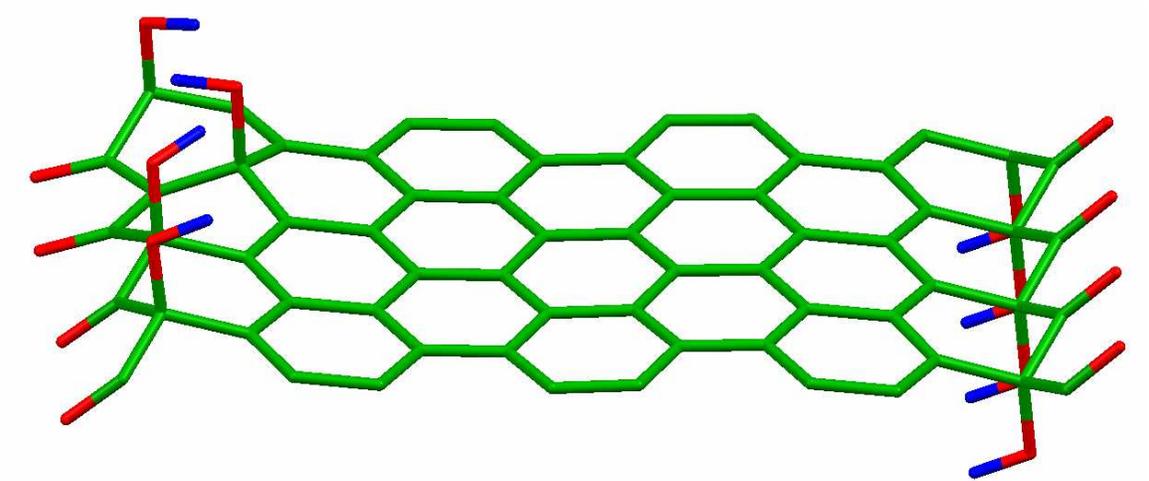

**Figure 6.** Optimized atomic structure for initial step of migration single hydroxyl group from the edge of oxidized GNR functionalized with hydroxyl groups to the centre.